\documentclass[epjST]{svjour}
\usepackage{graphics}
\usepackage{slashed,color}
\begin{document}
\title{Inverse magnetic catalysis - how much do we know about?}
\author{Aritra Bandyopadhyay\inst{1}\fnmsep\thanks{\email{aritra@m.scnu.edu.cn}} \and Ricardo L S Farias\inst{2}\fnmsep\thanks{\email{ricardo.farias@ufsm.br}} }
\institute{Guangdong Provincial Key Laboratory of Nuclear Science, Institute of Quantum Matter, South China Normal University, Guangzhou 510006, China \and Departamento de F\'{i}sica, Universidade Federal de Santa Maria, Santa Maria, 
 RS, 97105-900, Brazil }
\abstract{
Some of the advances made in the literature to understand the phase transitions of quark matter in the presence of strong magnetic field and finite temperature (zero quark chemical potential) are reviewed. We start by discussing the physics behind the Magnetic catalysis (MC) at zero/finite temperature and then focus on the lattice predictions for inverse magnetic catalysis (IMC) at high temperature and strong magnetic fields. Possible explanations for the IMC are covered as well. Finally, we discuss recent efforts to modify QCD (quantum chromodynamics) effective models in order to reproduce the IMC observed on the lattice simulations. We emphasize the fact that applying thermo-magnetic effects on the coupling constant of the NJL model significantly improve the effectiveness of the NJL model to obtain a reasonable physical description of hot and magnetized quark matter being in agreement with lattice results.} 
\maketitle
\section{Introduction}
\label{intro}

The existence of strong anisotropic magnetic fields has long been observed in the very early stages of non-central heavy ion collisions. Calculations based on classical electrodynamics was in agreement with this observation and showed that in the initial stages of the collision the value of the magnetic field perpendicular to the reaction plane reaches up to $10^{18}-10^{19}$ Gauss, exceeding the critical value of Schwinger field~\cite{bzdak,McLerran}. The parallel timeline between the formation of Quark Gluon Plasma after the heavy ion collision and the relaxation time of the generated magnetic field due to the collision subsequently indicates the phenomenological significance of the magnetic field and its influence on different aspects of the physics of heavy ion collisions. Several novel phenomena like the chiral magnetic effect (CME)~\cite{cme1,cme2,cme3} have been emerged subsequently as a result of taking the external magnetic field into consideration. The effect of the magnetic field on the QCD phase diagram has been studied vividly throughout the world, using lattice simulations and effective model calculations among others (For extensive reviews look at Refs~\cite{Miransky:2015ava,Andersen:2014xxa}). Two contrasting effects have grabbed the main attention in almost all of those studies : magnetic catalysis, which shows enhancement in the values of the quark condensate with increasing magnetic field (mostly at low temperature) and inverse magnetic catalysis, i.e. decreasing values of the condensate with increasing magnetic field (close to the transition temperature). Whereas the former is well explored and various studies have converged on its mechanism and theoretical basis, IMC appears as counterintuitive and somewhat puzzling. As we will see later in this mini-review, it is a result of an intricate competition, active only on certain temperature ranges and effective only for light quarks. The search for a complete theoretical understanding of the IMC effect is still ongoing and this gives us the proper platform to assess the current situation of the same through this mini-review.

This mini-review is organized as follows. In section~\ref{mcat} we discuss about the exploration of the magnetic catalysis effect. Thereafter, the lattice predictions about the inverse magnetic catalysis and subsequent advancements in light of lattice QCD have been discussed in section~\ref{latticeqcd}. In section~\ref{effectivemodels} we mainly focus on the progresses made by the effective QCD model calculations, more specifically the inclusion of effective thermo-magnetic effects on the coupling constant in the local Nambu--Jona-Lasinio (NJL) model and show corresponding results on multiple aspects of QCD. Finally we conclude in section~\ref{Conclusions}. 

\section{Magnetic catalysis}
\label{mcat}

The effect of an external electromagnetic field on the fermion mass was studied long ago by Klevansky and Lemmar~\cite{Klevansky:1989vi} in light of NJL model where the contrasting effects between  constant electric and magnetic fields were clearly demonstrated. It was closely followed by studies in the context of linear $\sigma$ model by Suganuma and Tatsumi~\cite{Suganuma:1990nn} which showed exactly similar effects. The external electric fields restored the chiral symmetry, whereas the magnetic fields were found to be acting as a positive catalyst of the chiral symmetry breaking. Similar effect of the constant external magnetic fields on the dynamical symmetry breaking was again confirmed by studies within the $(2+1)$ dimensional Gross-Neveu model, both at finite temperature~\cite{Klimenko:1991he,Klimenko:1992ch} and density~\cite{Krive:1992xh,Krive:1991uu}. This effect shown by the constant magnetic fields was puzzling in regard to most of the early model calculations, specially while drawing analogies with the completely opposing results from the Bardeen - Cooper - Schrieffer (BCS) theory of superconductivity. Klevansky first tried to clarify this issue in her review~\cite{Klevansky:1992qe} by pointing out the fact that the condensates in the NJL model consist of quark and antiquark pairs carrying opposite charges which is not the same as usual Cooper pairs consisting of the identical charge particles. Also about the contrasting behaviors of the external electric and magnetic fields it was quoted in Ref~\cite{Klevansky:1992qe} : `We can understand this physically, since we may imagine that the electric field destroys the condensate by pulling the pairs apart, while the magnetic field aids in anti-aligning the helicities which are bound by the NJL interaction'. 

Regardless of these developments still the theoretical basis of MC was incomplete. This was subsequently completed in Refs~\cite{Gusynin:1994re,Gusynin:1994va,Gusynin:1994xp,Gusynin:1995nb,Leung:1996qy,Lee:1997zj,Lee:1997uh} connecting the dynamical mass generation for the fermions with another well known fact - the dimensional reduction $D\rightarrow D-2$ of fermion dynamics in a constant anisotropic strong external magnetic field. It was known long ago that the motion of charged particles is quantized into several discreet degenerate energy levels in the perpendicular directions to the magnetic field, known as Landau levels. The energy gap between two Landau levels is dependent on the external magnetic field ($eB$) which means that at very high values of $eB$, the fermions can't jump beyond the lowest Landau level. Utilising both $(2+1)$~\cite{Gusynin:1994re,Gusynin:1994va} and $(3+1)$~\cite{Gusynin:1994xp,Gusynin:1995nb} dimensions within NJL model, Gusinyn, Miransky and Shovkovy illustrated that in the influence of a strong enough external magnetic field, the fact that the fermions are confined within the lowest Landau level effectively reduces the dynamics of fermion pairing by two spatial dimensions exposing them to much more severe infrared singularities. This catalyses the enhancement of the generation of a dynamical mass for the fermions even at the weakest attractive interaction between fermions and antifermions. The above statement can also be visualized by looking at the $(3+1)$ dispersion relation for fermions in a constant external magnetic field along the $Z$ direction (${\bf B} = B {\hat z}$), i.e. $E_n(k_z) = \sqrt{k_z^2+m_f^2+|q_fB|(2n+1-s)}$ where $n$ and $s$ respectively denotes the Landau level and the spin. This is similar to a dimensionally reduced $(1+1)$ dispersion relation with effective mass $M_f=\sqrt{m_f^2+|q_fB|(2n+1-s)}$.  In Ref~\cite{Gusynin:1995nb}, they have also argued that the dimensional reduction is indeed consistent with the spontaneous chiral symmetry breaking and Nambu-Goldstone (NG) modes do exist in presence of a strong external magnetic field. This argument was necessary to counter the Mermin-Wagner-Coleman (MWC) theorem~\cite{Mermin:1966fe,Coleman:1973ci} which says that gapless NG bosons cannot exist in dimensions less than $2+1$. Gusinyn, Miransky and Shovkovy argued that in the present problem, the condensate is a neutral excitation with respect to the external magnetic field and hence the dimensional reduction doesn't apply on them. As the discussion above suggests eventually this universal mechanism of dynamical mass generation led to the coining of the term magnetic catalysis which refers to either enhancement of an existing condensate or appearance of a new condensate by the presence of an external magnetic field. 

Numerous extensive further studies have found magnetic catalysis within NJL~\cite{Klimenko:1998su,Ghosh:2005rf,Menezes:2008qt,Menezes:2009uc,Boomsma:2009yk,Fayazbakhsh:2010bh,Chatterjee:2011ry,Gorbar:2000ku} and its Polyakov loop extended version~\cite{Fukushima:2010fe,Gatto:2010qs}, quark meson model~ \cite{Andersen:2011ip,Andersen:2012bq} and its Polyakov loop extended version~\cite{Mizher:2010zb}, chiral perturbation theory~ \cite{Shushpanov:1997sf,Agasian:2000hw,Andersen:2012zc}, renormalization group methods~ \cite{Hong:1996pv,Semenoff:1999xv,Fukushima:2012xw,Kamikado:2013pya,Kojo:2013uua,Andersen:2013swa}, analytic field theoretical approaches~ \cite{Gusynin:1999pq,Alexandre:2000yf,Miransky:2002rp,Ayala:2009fv,Ayala:2016awt}, Dyson-Schwinger equation~\cite{Mueller:2014tea}, holographic models~\cite{Zayakin:2008cy,Filev:2009xp,Filev:2010pm,Bolognesi:2011un,Alam:2012fw} and lattice QCD simulations~\cite{Buividovich:2008wf,DElia:2010abb,DElia:2011koc,Bali:2011qj,Bali:2012zg,Endrodi:2013cs} etc. 

\section{Inverse magnetic catalysis - advancements by Lattice QCD}
\label{latticeqcd}

From the last section it is evident that almost all the studies within magnetized medium led one to expect magnetic catalysis regardless of the temperature. But the situation took an unexpected turn when Lattice QCD calculations were made more precise, numerically. Let us discuss about that and the corresponding consequences in the present section.

 The early lattice QCD simulations~\cite{DElia:2010abb} for $N_f=2$ at finite temperature and magnetic fields had used larger quark masses vis-\'a-vis unphysical pion masses ($m_\pi$) within the range of 200 - 480 MeV. Moreover they were not continuum extrapolated, all of these resulting in signs of magnetic catalysis, i.e. an enhancement in the pseudocritical temperature ($T_{pc}$) with increasing magnetic field. Things changed when similar simulations for $N_f=2+1$ were carried out by Bali et al.~\cite{Bali:2011qj,Bali:2012zg}, but this time extrapolating the continuum and more importantly, using lower values for the quark mass corresponding to physical pion mass of $m_\pi = 140$ MeV. Quite surprisingly, their results show opposite signs from that of magnetic catalysis, a decreasing $T_{pc}$ with increasing magnetic field. Their studies also revealed that the magnetic field dependence of the condensates is complex, non-monotonic and a strong function of the temperature. Though when they increase the light quark masses up to the physical strange quark mass (i.e. $N_f=3$), results were quite different, showing increasing values of condensates with the magnetic field for all temperatures, similar to that of Ref~\cite{DElia:2010abb}. This behaviour led them to conclude that the change in $T_{pc}(B)$ is strongly affected by the difference in the used quark masses~\cite{Bali:2011qj}. For light quarks the indirect interaction between the magnetic field and the gauge degrees of freedom is enhanced, specially near the chiral limit, which results in a decreasing $T_{pc}$~\cite{Bali:2012zg}. 
 
To give a more detailed and proper explanation linking between these several contradicting results for the magnetic field dependence on the transition temperature, Bruckmann, Endr\"{o}di and Kovacs analyzed the lattice results in details~\cite{Bruckmann:2013oba} separating the sea and valence contributions to the condensate, a method previously introduced by D'Elia and Negro~\cite{DElia:2011koc}. We discuss that below. 

The chiral condensate in presence of an external magnetic field can be defined as 
\begin{eqnarray}
\langle\bar\psi\psi\rangle (eB) = \frac{1}{\mathcal{Z}(eB)}\int dU e^{-S_g}{\rm det}[\slashed{D}(eB)+m]{\rm Tr}[\slashed{D}(eB)+m]^{-1},
\end{eqnarray}
where the partition function $\mathcal{Z}(eB)$ is given as
\begin{eqnarray}
\mathcal{Z}(eB)= \int dU e^{-S_g}{\rm det}[\slashed{D}(eB)+m],
\end{eqnarray}
and $S_g$ is the pure glue action and $U$ are the gauge configurations. It is evident from the above expression that the magnetic field enters in the chiral condensate through both the measure, i.e. the functional determinant and the trace of the propagator. These dependences can be separated out by defining the so-called valence and sea condensates, i.e. 
\begin{eqnarray}
\langle\bar\psi\psi\rangle_{\rm val}(eB) &=& \frac{1}{\mathcal{Z}(0)}\int dU e^{-S_g}{\rm det}[\slashed{D}(0)+m]{\rm Tr}[\slashed{D}(eB)+m]^{-1}, \nonumber\\
\langle\bar\psi\psi\rangle_{\rm sea}(eB) &=& \frac{1}{\mathcal{Z}(eB)}\int dU e^{-S_g}{\rm det}[\slashed{D}(eB)+m]{\rm Tr}[\slashed{D}(0)+m]^{-1}.\nonumber
\end{eqnarray}
So the valence condensate is where the measure of the gauge configurations $U$ is done at vanishing magnetic field and the sole magnetic field contribution comes from the trace of the propagator. The sea contribution is the average of the trace of the propagator taken in zero magnetic field, but where the measure is done at nonzero B. 
 Because in QCD all the fermions are electrically charged, it seems difficult to separate the valence and the sea effects of the magnetic field on the condensates. But it can be done by using similar techniques as used in partially quenched QCD~\cite{Morel:1987xk,Sharpe:2006pu,Verbaarschot:2000dy}. The idea is to cancel the redundant part of the functional determinant (for both valence and sea contributions) by generating inverse determinants using ghost quarks. Though the formulation of lattice QCD naturally separates the valence and the sea effects by the virtue of the dynamical gauge fields, but as a downside an apparent gauge dependence appears in the process.
To explicitly see the effects of the valence and the sea contrbutions, in Ref~\cite{Bruckmann:2013oba} the authors observed the magnetic field dependence of the difference $\Delta\Sigma(B) \propto (\langle\bar\psi\psi\rangle_B -\langle\bar\psi\psi\rangle_0 )$ for both valence and sea contributions for the up quark at different temperatures chosen to be close to and well-below $T_{pc}$. It is seen that though the valence contribution increases the condensate for both cases, the sea contribution enhances the condensate only well-below $T_{pc}$ and suppresses it around $T_{pc}$. Eventually, very close to $T_{pc}$, i.e. around $T=160$ MeV, the suppression of the condensates through sea contribution dominates the enhancement of the same through valence contribution. This results in a net suppression of the condensates, termed as Inverse Magnetic Catalysis~\cite{Bali:2012zg,Bruckmann:2013oba}.  

Furthermore, the isolated effect on the sea contribution was carefully analyzed by Bruckmann et al. in Ref~\cite{Bruckmann:2013oba}.  The magnetic field dependent fermion functional determinant results in a reweighting in $B$, i.e. $\langle \mathcal{O} \rangle\rightarrow \langle\mathcal{O}\rangle_{eB}$ where in this case $\mathcal{O} \equiv {\rm Tr}[\slashed{D}+m]^{-1}$ and the resulting expectation value can be expressed as 
\begin{eqnarray}
\langle\mathcal{O}\rangle_{eB} = \langle e^{-\Delta S_f(eB)} \mathcal{O}\rangle_0 / \langle e^{-\Delta S_f(eB)} \rangle_0.
\end{eqnarray}
The reweighting factor $\Delta S_f(eB)$ is defined as the change in the fermionic action
 \begin{eqnarray}
 -\Delta S_f(eB) = \log{\rm det}(\slashed{D}(eB)+m) -  \log{\rm det}(\slashed{D}(0)+m).
 \end{eqnarray}
 whose fluctuations can subsequently affect the condensates, because of the correlation between them. Larger values of condensates correspond to larger fermionic actions and hence they are suppressed more by the functional determinant. This effect, known as the `sea' effect has been identified as the main reason behind IMC~\cite{Bruckmann:2013oba}. On the other hand in the model calculations there are usually no dynamical gauge fields present and hence sea effect can't be observed there. This particular failure of model calculations in generating the sea effect and hence predicting IMC around transition temperatures for light quarks has motivated a large body of work attempting to clarify the reasons for the observed discrepancies~\cite{fragaMIT,prl,leticiaLN,EPNJL3,debora,bruno,Tawfik:2014hwa,Tawfik:2016lih,Tawfik:2017cdx,Tawfik:2019rdd,Farias:2015eea,ayala1,Mueller:2015fka,ferreira1,ayala2,ayala3,ferrer1,cao1,sadooghi1,andersen2,kanazawa,ayala4,PhysRevD.92.096011,Ayala:2015qwa,huang1,Hattori1,fengli1,raya,prdferreira,ayala7,prdmarcus,jpcsGB0,mao,Farias:2017svg,Farias:2019wdh,Mao:2017tcf,GomezDumm:2017iex,Ballon-Bayona:2017dvv,Pagura:2016pwr,Li:2016gfn,Ayala:2018wux,Endrodi:2019whh,Endrodi:2019zrl,ayalafit,Mukherjee:2018ebw,Feng:2014bpa,Feng:2015qpi,Chaudhuri:2019lbw}. In~\cite{massimo} a very interesting 
 lattice QCD simulation were implemented for $N_f =2+1$ QCD in a magnetic background considering the whole range of the pion masses. Their results claim that the modifications induced by the magnetic background on the gauge field distribution and on the confining properties of the medium could play a primary role, and that the idea of inverse magnetic catalysis might be a secondary phenomenon. In the next section we will focus on a particular approach taken by local NJL model calculations and its effectiveness by showing corresponding findings. 
 
\section{Inverse magnetic catalysis in effective models of QCD}
\label{effectivemodels}

The surprising and unexpected IMC effect was not predicted by any effective models of QCD in the literature, and the idea was founded based on the lattice simulations considering physical pion mass $m_{\pi}=135$ MeV, which corresponds to 
the regime of small current quark masses. Indeed, an explanation was needed about why IMC was not observed in effective models. There are numerous proposals and alternative mechanisms trying to explain IMC, as discussed/mentioned before. In this section we will revisit the recent progress we have done including IMC effects on the description of the hot and magnetized quark matter using effective models of QCD. In particular, we will focus in the local version of the Nambu-Jona--Lasinio model~\cite{njl}, including thermo-magnetic effects on the coupling constant of the model in such a way that describes the lattice data. 

\subsection{Thermo-magnetic effects in the NJL coupling constant}
\label{gBTprc+epja}

The first work on the literature to obtain IMC in the local version of NJL model was done in~\cite{prc14}. In this work the authors pointed out that the apparent failure of effective models in providing IMC can be explained by recalling that the couplings in those models do not run with the magnetic field, in contrast to QCD whose coupling, as Miransky and Shovkovy have shown~\cite{igor}, decreases for large B.

Ref.~\cite{Bali:2012zg} presents lattice results for the average $(\Sigma_u + \Sigma_d)/2$ 
and the difference $\Sigma_u - \Sigma_d$, where $\Sigma_f = \Sigma_f(eB,T)$ is defined as
 \begin{equation}
\Sigma_{f}(eB,T) = \frac{2m_{f}}{m_\pi^2 f_\pi^2}\left[\langle \bar \psi_f\psi_f \rangle 
- \langle \bar \psi_f\psi_f \rangle_0\right]+1 ,
\label{sigmalatt}
 \end{equation}
with $ \langle \bar \psi_f\psi_f \rangle_0$ being the quark condensate at $T=0$ and $eB=0$.

Therefore, in~\cite{prc14} the authors introduced an ansatz that makes the local NJL model coupling constant G a function of $eB$ and $T$, in a similar way as the strong-coupling runs in QCD. As a result, they showed that, at $T = 0$, the coupling constant can be given as

\begin{eqnarray}
G(eB) =\frac{G_0}{1+ \alpha \ln \left(1 + \beta \, \frac{eB}{\Lambda^2_{QCD}}\right)},
\label{run_G}
\end{eqnarray}

\noindent with $G_0 = 5.022$~{\rm{GeV}$^{-2}$}, which is the value of the coupling at $eB=0$. The free parameters $\alpha$ and $\beta$ were fixed to obtain a reasonable 
description of the lattice average $(\Sigma_u + \Sigma_d)/2$ for $T=0$.

Working at finite temperature the authors similarly suggested a coupling constant $G(eB,T)$ given by
\begin{equation}
G(eB,T) = G(eB) \, \left(1 - \gamma\, \frac{|eB|}{\Lambda^2_{QCD}}\,
\frac{T}{\Lambda_{QCD}}\right),
\label{GT}
\end{equation}
Where $\gamma$ is another parameter fitted to reproduce the lattice results for the average of the condensates, given in Ref.~\cite{Bali:2012zg}.

In Ref.~\cite{prc14} the authors worked in the physical limit, i.e. with nonzero current quark masses, so that at high temperatures the model displays a crossover where chiral symmetry is partially restored. They chose the vacuum normalized quark condensates as an observable to define the pseudocritical temperature $T_{pc}$ and defined the thermal susceptibilities $\chi_T$ as  

\begin{equation}
 \chi_T=-m_{\pi}\frac{\partial \sigma}{\partial T} \,\,,
\label{chiT}
\end{equation}

\noindent to determine the behavior of $T_{pc}$ with the magnetic field. Furthermore, the ratio $\sigma$ was defined as
\begin{equation}
\sigma=\frac{\langle\bar{\psi}_u\psi_u\rangle(eB,T)+\langle\bar{\psi}_d\psi_d\rangle(eB,T)}
{\langle\bar{\psi}_u\psi_u\rangle(eB,0)
+\langle\bar{\psi}_d\psi_d\rangle(eB,0)}\,\,.
\end{equation}

\begin{figure}
\begin{center}
\resizebox{0.55\columnwidth}{!}{
  \includegraphics{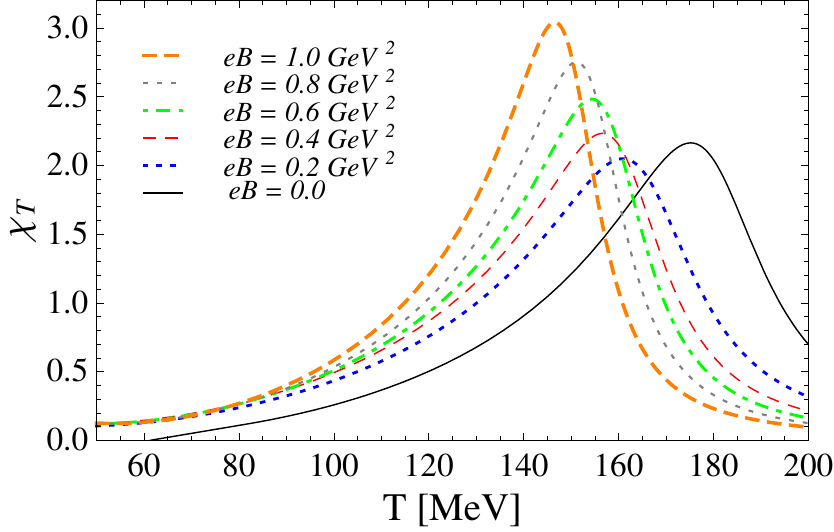}}
  \end{center}
\caption{The normalized chiral susceptibility evaluated in NJL considering $G(eB,T)$ as a function of the temperature for different values of the magnetic field eB. Figure taken from~\cite{prc14}.}
\label{fig:1}  
\end{figure}

The behavior of $\chi_T$ as a function of the temperature is shown in Fig.~\ref{fig:1} and the peak defines the pseudo critical temperature for chiral symmetry restoration. The peak shifts to the left when the magnetic field is increased indicating that the NJL model with $G(eB,T)$ presents IMC effects at high temperature. 

In Ref.~\cite{epja} the authors improved the ansatz used in Ref.~\cite{prc14}, in order to get a better quantitative agreement with the lattice results. They have shown that this improved model can be used in a plethora of situations in order to analyze the behavior of important physical quantities such as thermodynamics and magnetization.

\begin{figure}
\begin{center}
\resizebox{0.55\columnwidth}{!}{
  \includegraphics{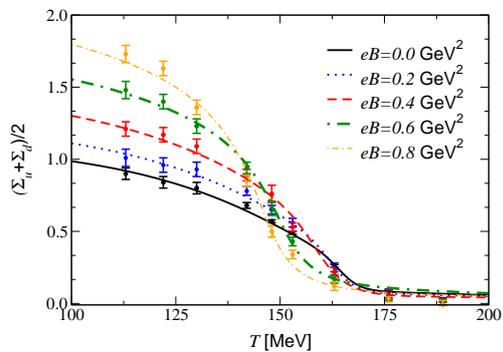}}
  \end{center}
\caption{Condensate average as a function
of temperature for different values of the magnetic field for 
$G(eB,T)$. Data points from Ref.~\cite{Bali:2012zg} and figure taken from~\cite{epja}.}
\label{fig:2}       
\end{figure}

Their parameters were fixed in order to reproduce the lattice results of Ref.~\cite{Bali:2012zg} for the quark condensate average, $ (\Sigma_u + \Sigma_d)/2$, and the following interpolation formula was used for the NJL coupling constant:
\begin{eqnarray}
G(eB,T) = c(eB)\left[1-\frac{1}{1+ e^{\beta(eB) \left[T_a(eB) - T\right]}}\right]+s(eB) .
\label{ourGBT}
\end{eqnarray}
In Table~\ref{glatnjl} we can see the values of the parameters $c,s,\beta$ and $T_a$, which depend only 
on the magnetic field. The choice for this expression was arbitrary, any other form that fits the lattice data is expected to give the same qualitative results for thermodynamical quantities in the appropriate $B$ and $T$ range. As we can see in Fig.~\ref{fig:2}, the NJL predictions with $G(eB,T)$ for the quark condensate average is numerically very close to those obtained within the LQCD simulations. 

\begin{table}
\caption{Values of the fitting parameters in Eq.~(\ref{ourGBT}). Units are in appropriate 
powers of GeV. Table taken from~\cite{epja}}
\label{glatnjl}
\begin{center}
\begin{tabular}{ccccc}
\hline\noalign{\smallskip}
$eB$ & $c$ &  $T_a$ & $s$   & $\beta$  \\ \hline\noalign{\smallskip}
0.0    &    0.900 &  0.168   &  3.731   &  40.000 \\\noalign{\smallskip}
0.2    &    1.226  &  0.168   &  3.262   &  34.117 \\\noalign{\smallskip}
0.4    &    1.769  &  0.169   &  2.294   &  22.988 \\\noalign{\smallskip}
0.6    &    0.741  &  0.156   &  2.864   &  14.401 \\\noalign{\smallskip}
0.8    &    1.289  &  0.158   &  1.804   &  11.506 \\\noalign{\smallskip}
\hline
\end{tabular}
\end{center}
\end{table}

\begin{figure}
\begin{center}
\resizebox{1.05\columnwidth}{!}{
  \includegraphics{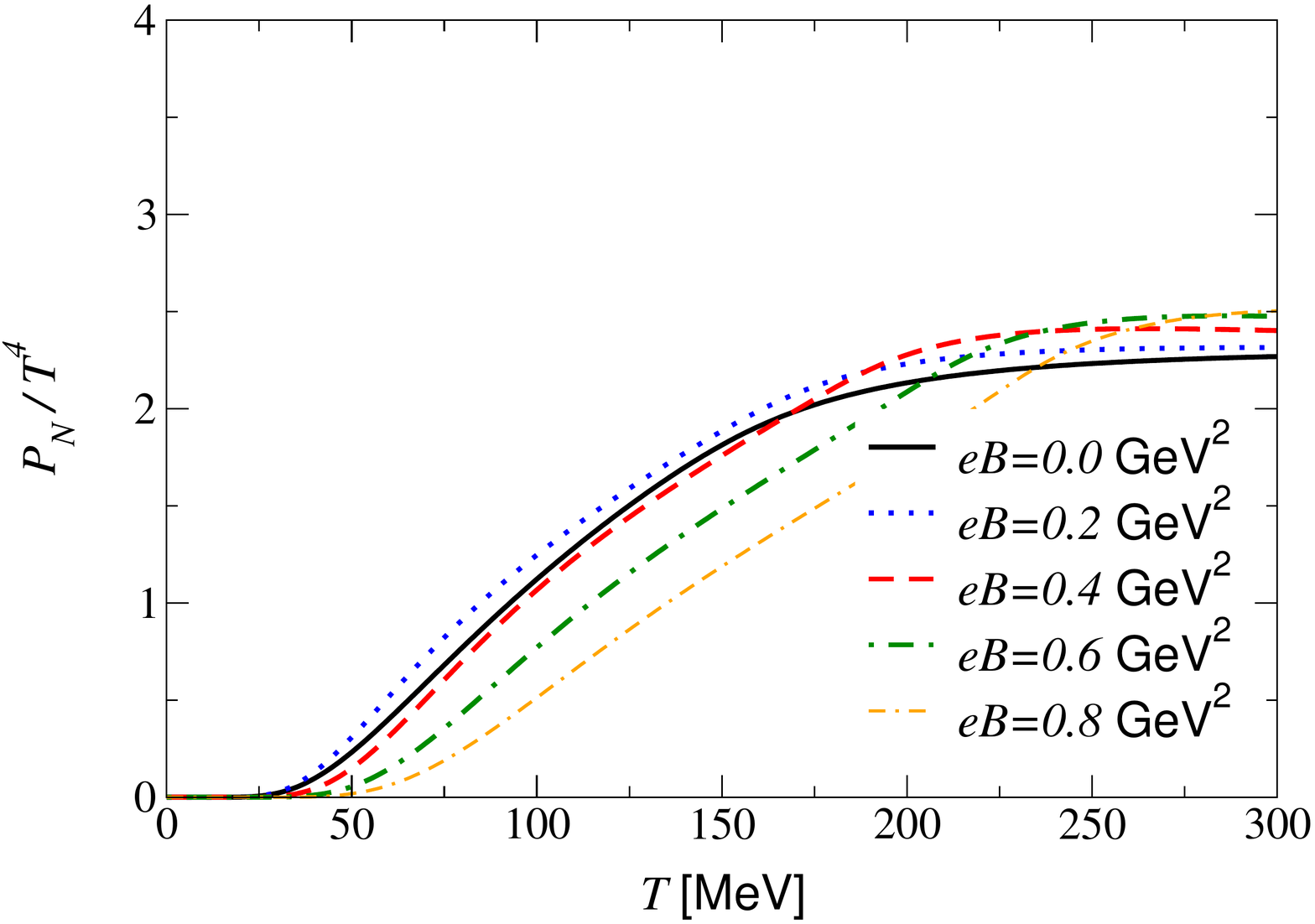}
  \includegraphics{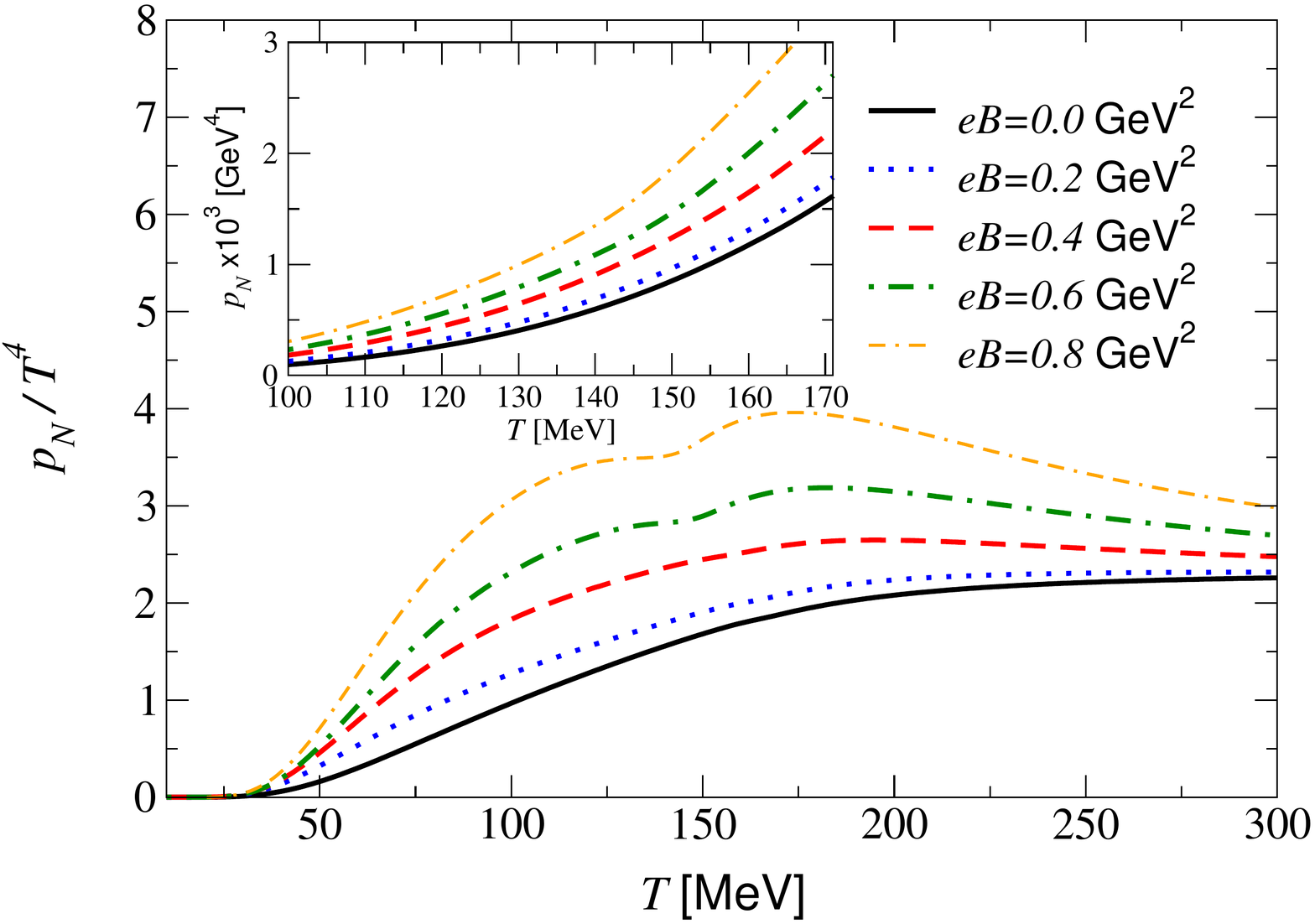}} 
  \end{center}
\caption{Normalized pressure as a function of temperature for different values of the magnetic field
                calculated with $G(0,0)$ (left) and $G(eB,T)$ (right). The inset in the right panel shows that, for a given $T$, the pressure always increases with~$eB$. Figures taken from~\cite{epja}.}
\label{fig:3}    
\end{figure}

With this expression of $G(eB,T)$ the authors have studied the QCD thermodynamics in NJL model and compared them with the usual NJL model with fixed coupling constant~\cite{epja}. Significant improvements were observed while comparing with the new NJL results including the thermo-magnetic effects with lattice results of Ref.~\cite{bali}. We can see in Fig.~\ref{fig:3} that the behavior of the pressure as a function of the temperature for different values of magnetic fields is strongly affected by the inclusion of the thermo-magnetic effects on the coupling constant of the model. The NJL results of Ref.~\cite{epja} using $G(eB,T)$ also show that the chiral transition becomes sharper and peaks observed in thermodynamic quantities increase considerably with $eB$, a feature consistent with the lattice simulations which is often missed while using NJL model with a fixed coupling constant.

\begin{figure}
\begin{center}
\resizebox{1.05\columnwidth}{!}{
  \includegraphics{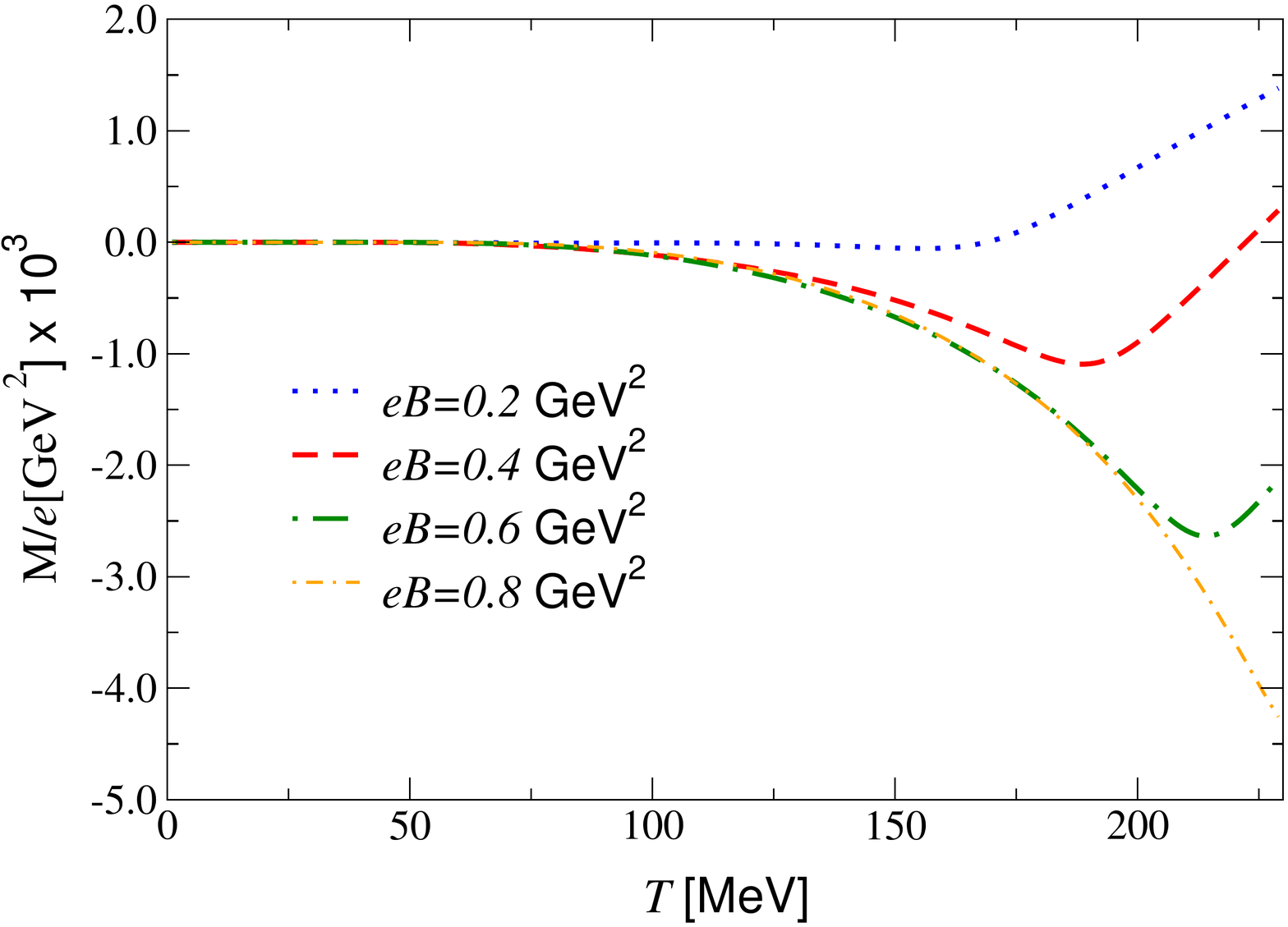}
  \includegraphics{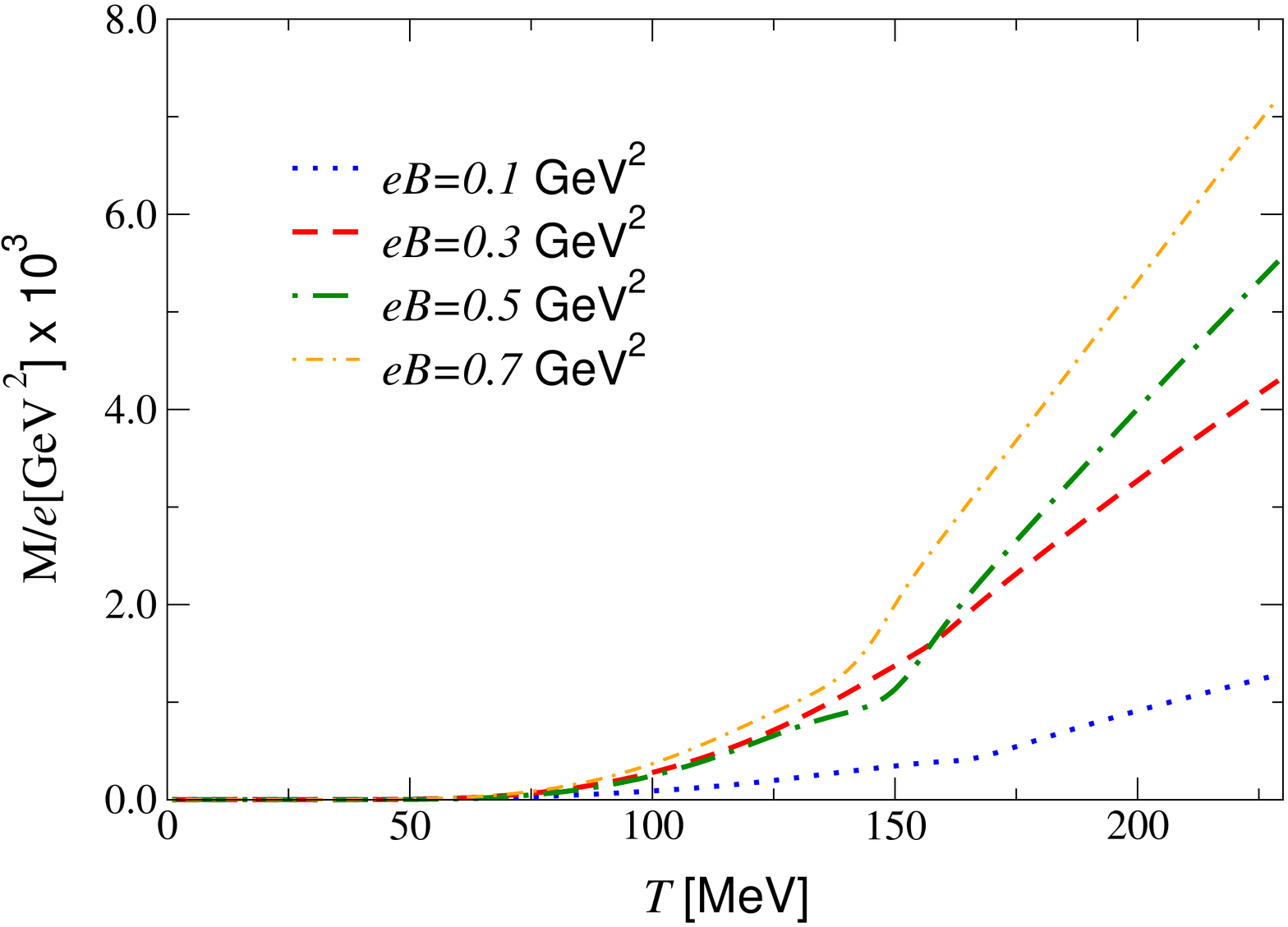}} 
  \end{center}
\caption{The normalized magnetization $M/e$ of quark matter as a function of the temperature for different values of the magnetic field strength obtained with $G(0,0)$ (left) and $G(eB,T)$ (right). Figures taken from~\cite{epja}.}
\label{fig:4}     
\end{figure}

These results seem to indicate that $eB$ and $T$ dependence in $G$, which gives the correct $T_{pc}$, is neither fortuitous nor valid for a single physical quantity; it seems to correctly capture the physics, left out in the conventional NJL model. It was verified again when the NJL model with thermo-magnetic effects on the coupling constant predicted that the magnetization is positive in all the temperature range, which complies to the paramagnetic nature of QCD medium and is in agreement with lattice calculations, as can be seen in Fig.~\ref{fig:4}. 

One may see that the scenario including $G(eB,T)$ seems to have a rather important feature which conciliates the results obtained within the NJL model and the lattice simulations. More details regarding this can be found in Ref.~\cite{epja}. Finally we would like to add that similar to the coupling constant, the thermomagnetic properties of the dynamical quark mass entering the NJL model are studied in Refs~\cite{ayalafit,Ayala:2017thy} as a key factor to understand the IMC phenomenon. 

\subsection{IMC on the neutral pion mass in a hot and magnetized matter}
\label{mpi0gBT}
Recently, much attention and efforts have been devoted to study and describe the importance of using an appropriate regularization scheme within NJL model to describe magnetized quark matter~\cite{norberto,ricardo}. Authors have explicitly shown that inappropriate regularization schemes may give rise to spurious solutions and nonphysical oscillations of the physical quantities~\cite{difreg}. One of these nonphysical solutions that needs to be circumvented is the imaginary meson masses~\cite{iranianos} which appears due to the inappropriate choice of the 
regularization procedure. In Ref.~\cite{mpi0TB} the authors have introduced magnetic field independent regularization scheme (zMFIR) based on the Hurwitz-Riemann Zeta function for the hot and magnetized medium within SU(2) NJL model, and calculated the neutral meson masses.

\begin{figure}
\begin{center}
\resizebox{1.05\columnwidth}{!}{%
  \includegraphics{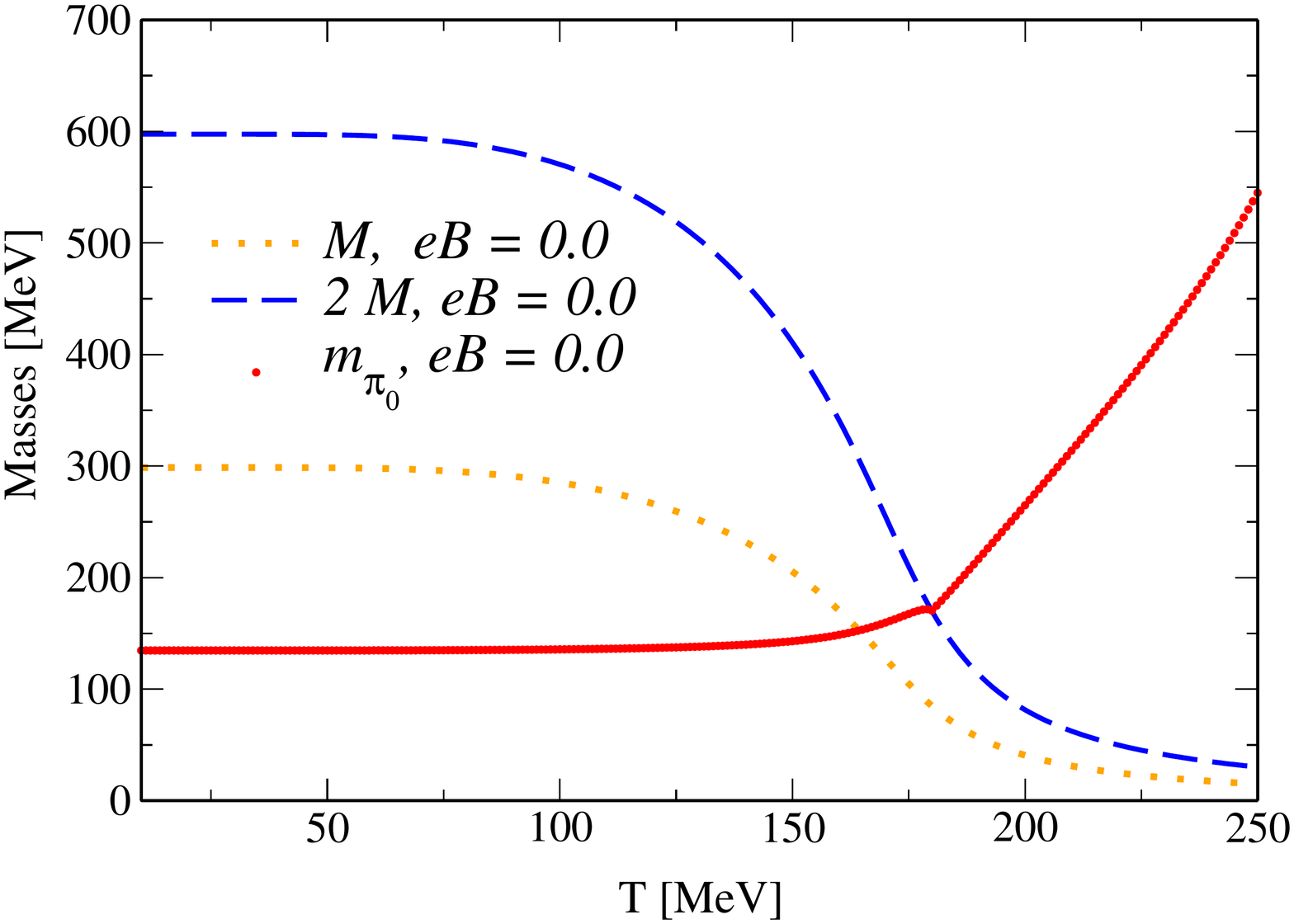}
  \includegraphics{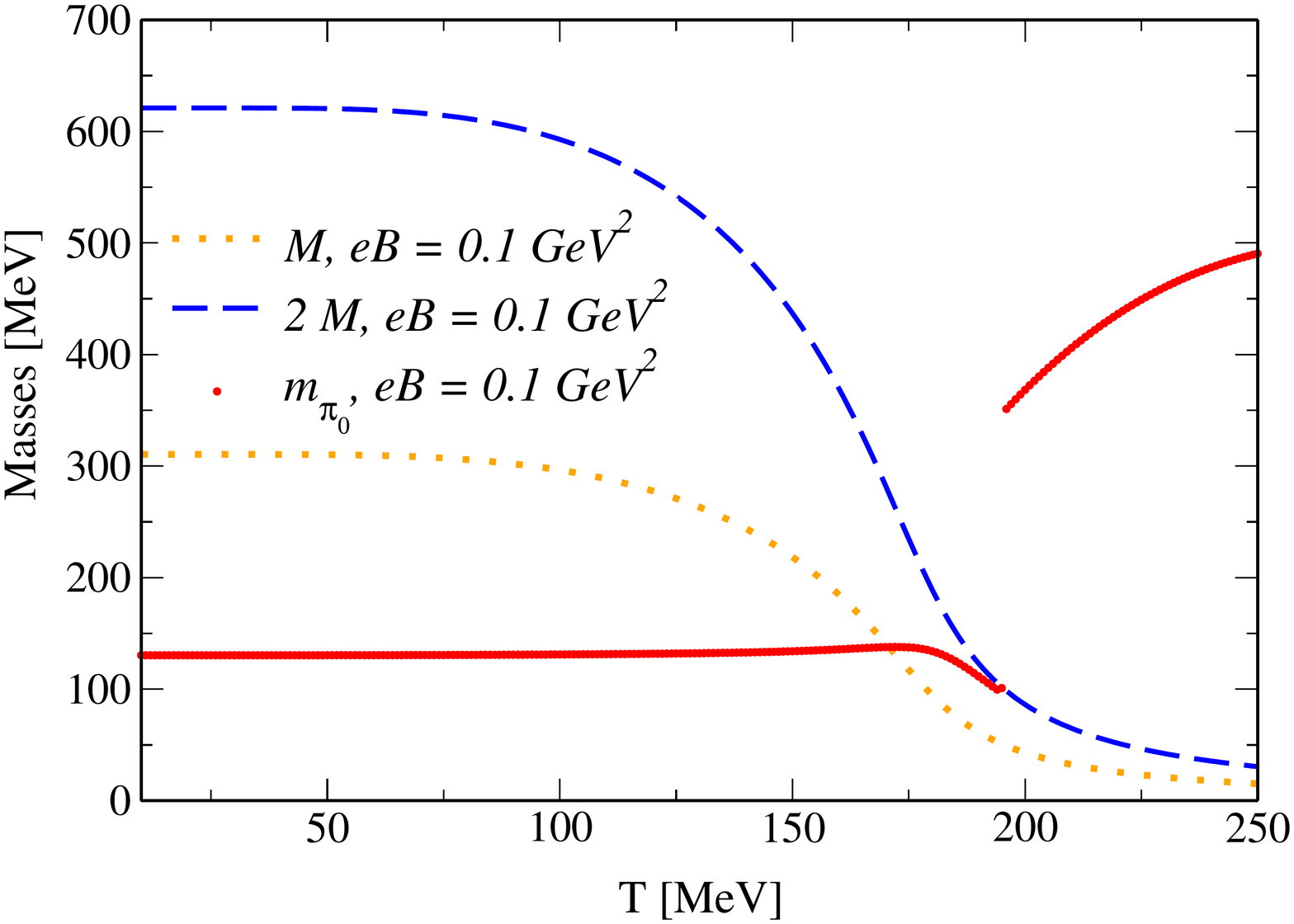}} 
  \end{center}
\caption{Masses as a function of the temperature for different values of magnetic field. Figures taken from~\cite{mpi0TB}.}
\label{fig:5}       
\end{figure}

In Fig.\ref{fig:5} we show the results obtained in Ref.~\cite{mpi0TB} using zMFIR for the behavior of the effective quark masses at finite temperature with $eB=0.0$ (left panel) and $eB=0.1$ GeV$^2$ (right panel). In the low temperature regime we see that the magnetic field enhances the breaking of chiral symmetry, which is a clear manifestation of the magnetic catalysis~\cite{Miransky:2015ava}. When the 
temperature is larger than the pseudo-critical temperature, the chiral symmetry restores partially which reduces the effective quark masses. It is the well-known scenario of MC where $T_{pc}$ increases with $eB$~\cite{prc14,epja}.

For low temperatures ($T<T_{pc}$) the mesons are protected by the Goldstone phase, but when we have 
partial restoration of chiral symmetry the neutral mesons enters in the Wigner-Weyl phase, where $\pi_0$ 
meson has a thermal excitation with a finite decay width. As we increase the magnitude of the magnetic field
the thermal excitations become more energetic when compared with the zero 
magnetic field case. We can see in Fig.~\ref{fig:5} that $\pi^0$ mass suddenly jumps to a more energetic solution at the dissociation temperature in presence of a magnetic field. The Mott dissociation temperature is defined when the system reaches $m_{\pi}(T)=2M(T)$. These more energetic resonances that authors obtained as they increased the magnetic field are direct results from the dimensional reduction of 
the system at strong magnetic fields which enforces the system to go to another state, since we have less states for the creation of the thermal $q-\bar{q}$ excitation. More details can be found in Ref.~\cite{mpi0TB}.

As we have discussed before, IMC is a phenomena that happens at high temperatures and for strong magnetic fields. To obtain a physical description of the behavior of neutral pion mass in a hot and magnetized medium, one needs to include IMC effect. We have shown that an efficient way to conciliate the NJL model results with lattice predictions for IMC effect is by the inclusion of thermo-magnetic effects on the coupling constant $G(eB,T)$ of the model~\cite{epja}. 
Some important results of Ref.~\cite{mpi0TB} are summarized in Table~\ref{Motttab}, where the authors have obtained that, at fixed coupling constant $G(0,0)$, $T_{Mott}$ is catalysed if the magnetic field is increased. They have worked in mean field approximation, which reinforces the observation of a catalyzed Mott Temperature. Similarly again, when they have included IMC effects on the model they have obtained an inverse magnetic catalysis on the Mott Temperature. 

\begin{table}
\caption{Values of the Mott dissociation temperature using fixed $G$ and $G(eB,T)$.}
\label{Motttab}
\begin{center}
\begin{tabular}{ccc}
\hline\noalign{\smallskip} & $T_{Mott}^{G(0,0)}$ [MeV] & $T_{Mott}^{G(eB,T)}$ [MeV]\\
$eB$ [GeV$^2$] &  &   \\ \hline\noalign{\smallskip}
0.1    &    195.0  &  166.84   \\\noalign{\smallskip}
0.2    &    200.2  &  164.90   \\\noalign{\smallskip}
\hline
\end{tabular}
\end{center}
\end{table}

\begin{figure}
\begin{center}
\resizebox{0.55\columnwidth}{!}{
  \includegraphics{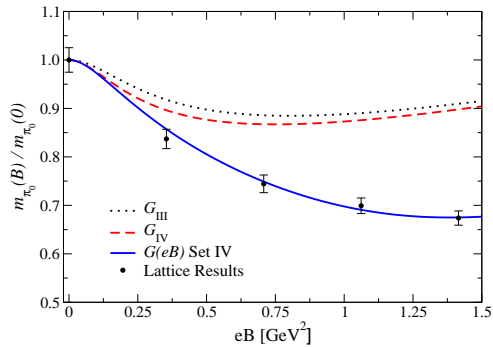}} 
  \end{center}
\caption{Normalized neutral pion mass $m_{\pi_0}(eB)/m_{\pi_0}(0)$ in the NJL model with different coupling schemes and a large current quark mass compared to recent lattice results~\cite{lattpion0}. Figure taken from~\cite{plb}.}
\label{fig:6}    
\end{figure}

The findings of Ref.~\cite{mpi0TB} with zMFIR scheme show that one can avoid unphysical results and proper choice of regularization scheme provides results in agreement with the lattice QCD results. In Ref.~\cite{plb} we can see another very interesting result that the authors have obtained by introducing magnetic effects on the NJL coupling constant, following the ansatz of Ref.~\cite{epja} at zero temperature $G(eB,0)$. They have evaluated the neutral pion mass as a function of the magnetic field and we can see in Fig.~\ref{fig:6} that the quantitative agreement between their results and recent LQCD predictions is remarkable. This in turn allows us to improve the applicability of NJL model on hot and magnetized quark matter.

\subsection{Axion in a hot and magnetized medium}
\label{axion}

Motivated by the very interesting results presented before with the improvement that the authors have done conciliating between NJL model predictions and LQCD results related to inverse magnetic catalysis, various other studies start to apply this approach on different aspects of the magnetized quark matter. One of the possible candidates for cold dark matter is the pseudo-Goldstone boson of a spontaneously broken global Abelian symmetry called Axion, that can be considered a solution for the absence of the charge and parity ($CP$) violation effects in QCD. In Ref.~\cite{axion} for the first time in the literature, to the best of our knowledge, the thermodynamical properties of QCD axions were studied within the NJL model in a hot and magnetized medium. 

\begin{figure}
\begin{center}
\resizebox{1.02\columnwidth}{!}{
  \includegraphics{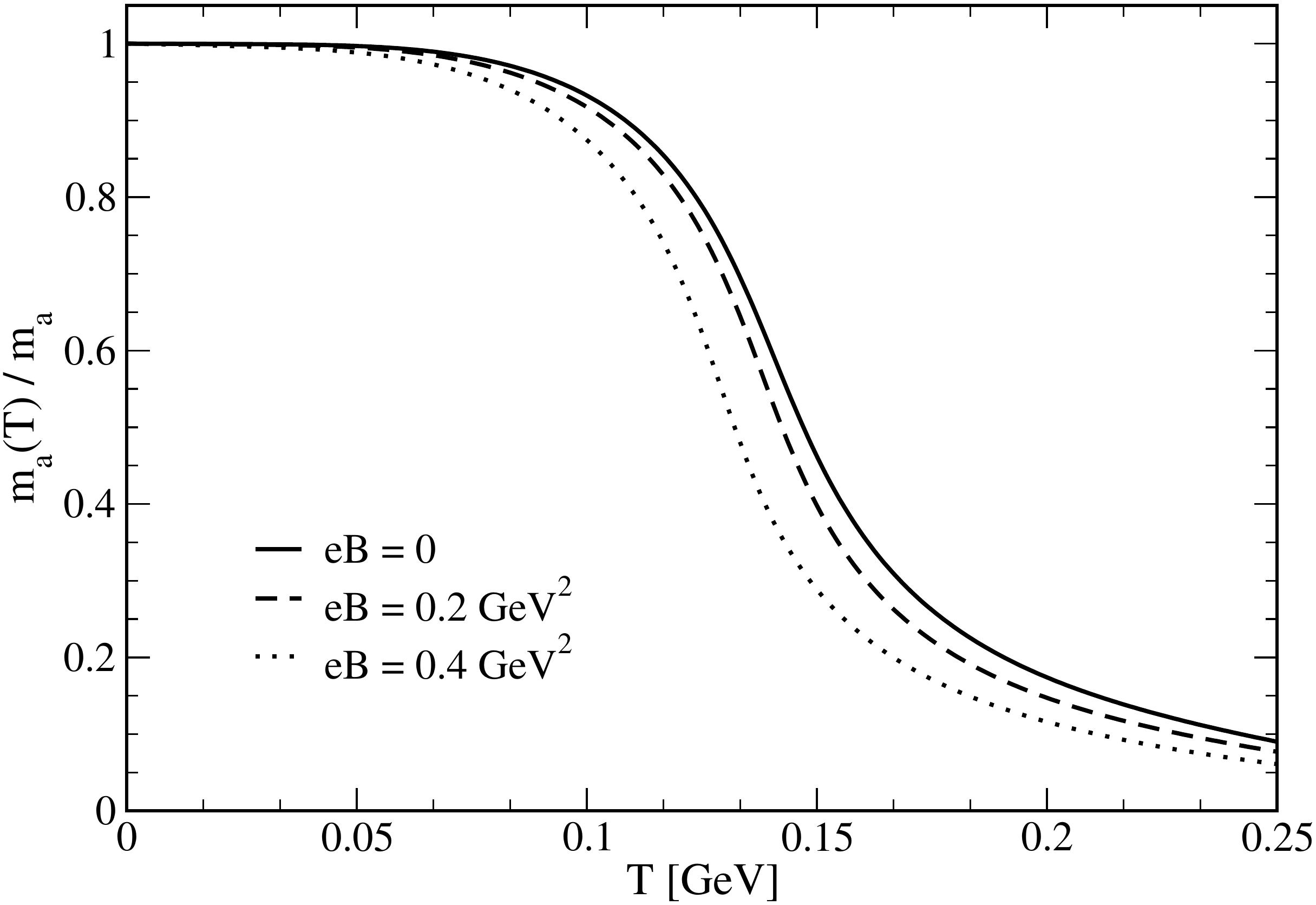}
    \includegraphics{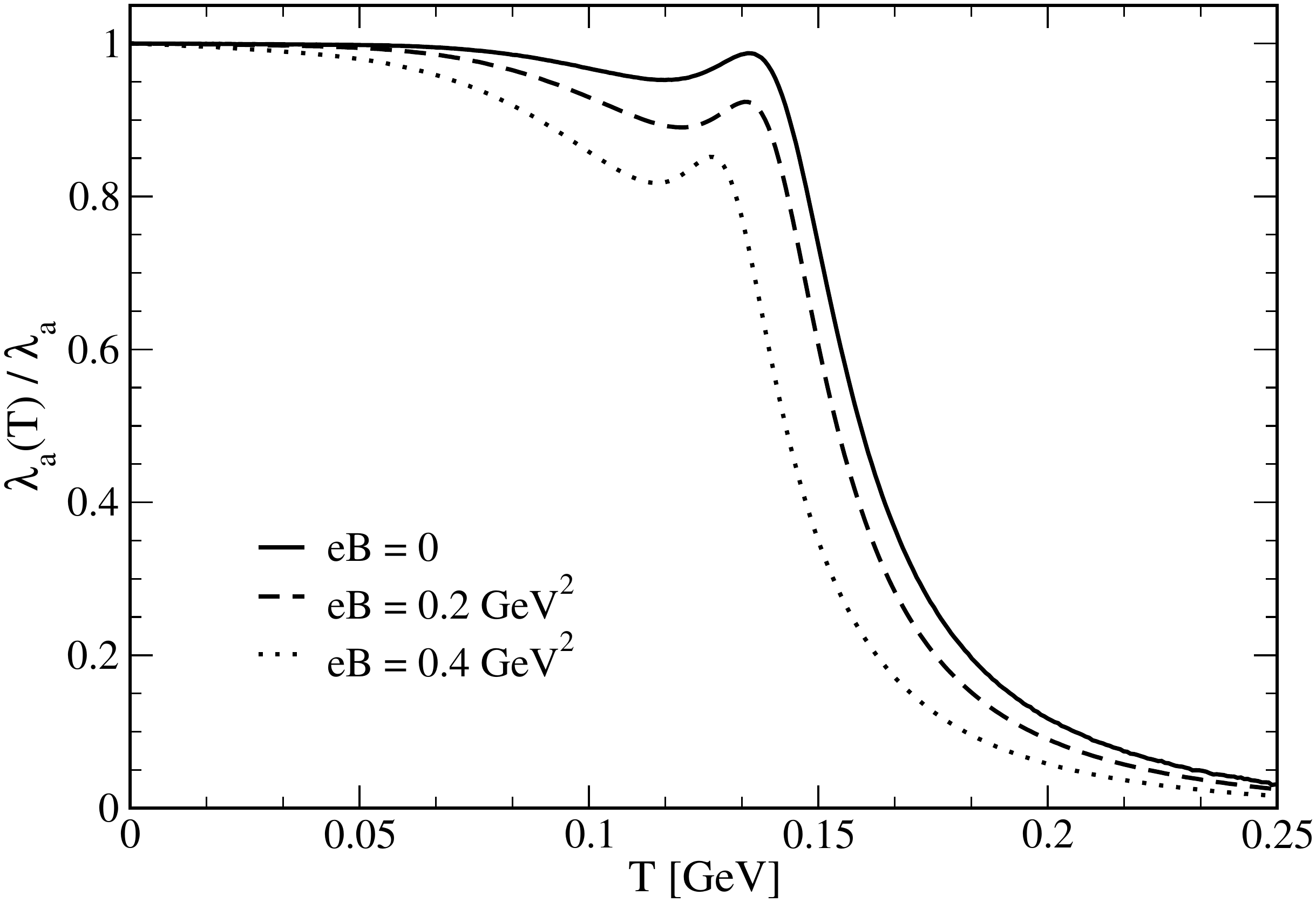}}
    \resizebox{0.52\columnwidth}{!}{
  \includegraphics{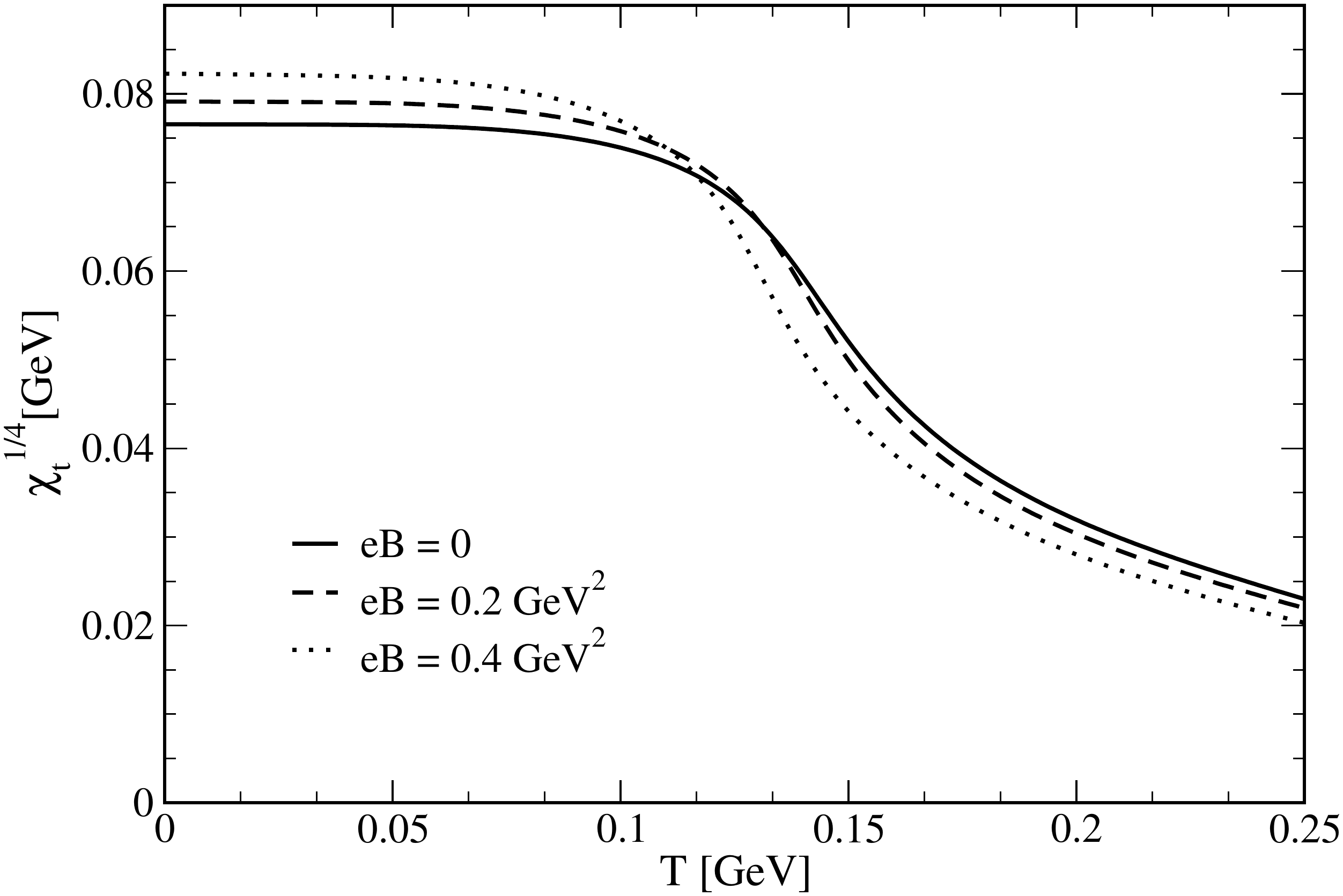}} 
  \end{center}
\caption{Variation of the axion mass ratio $m_a(T)/m_a$, the axion self-coupling
  ratio $\lambda_a(T)/\lambda_a$ and the topological
  susceptibility with $T$ using  $G(eB,T)$ for different values of $eB$. Figures taken from~\cite{axion}.}
\label{fig:7}   
\end{figure}

In {}Fig.~\ref{fig:7} we have shown the MC and IMC effects on the the variation of scaled axion mass, self-coupling and topological susceptibility with $T$ for the fully magnetic field and temperature dependent coupling constant $G(eB,T)$ (following the procedures adopted in Ref.~\cite{epja}). The IMC effect is noticeable around $T_{pc}$ in scaled axion mass, self-coupling because the inflection points, indicating the pseudo critical temperature location, is shifted towards lower temperature with increasing magnetic fields. The kink-like feature shown for the case of scaled axion self-coupling around $T_{pc}$ appears even for $eB=0$ case~\cite{Lu:2018ukl} and shows a similar shift towards lower temperature for increasing magnetic fields.  For the topological susceptibility we can clearly see both the MC and the IMC effects in different temperature regimes with a crossover happening in between them. 

In Ref.~\cite{axion} authors observed behaviors consistent with the well known Vafa-Witten theorem and Dashen's phenomena, where they have discussed the occurrence of different possible phases in details. They have studied the thermo-magnetic effects on the spontaneous $CP$ violation, thereby showing the magnetic field dependence of the critical temperature for the spontaneous $CP$ symmetry restoration. The authors have explored the phase diagram where the $CP$ phase transition and chiral crossover can be identified and the importance of including the thermo-magnetic effects on the NJL coupling constant in the case of a magnetized quark matter. 

The quantities such as the axion mass, self-coupling and topological susceptibility are very important regarding the study of cold dark matter, axion stars, the cooling anomaly problem in astrophysics. In Ref.~\cite{axion} the authors showed the importance of the IMC to obtain a good physical description of these quantities, specially in the critical region where we have  $CP$ phase transition and chiral symmetry partially restored.

\section{Final Remarks}
\label{Conclusions}

In this mini review we have  talked about two of the most important observations which came out from the exploration of the QCD phase diagram in presence of the magnetic field, i.e. MC and IMC. The mechanism of MC has been long discovered which makes the theoretical understanding behind it complete. But the same can not be said for IMC despite of several past and presently ongoing studies. Lattice QCD results detected that this counterintuitive behavior is actually arising because of the presence of the dynamical gauge fields which are inherently absent in model calculations. But over the years model calculations have developed some effective methods of capturing the IMC effect in their results, one of which is the inclusion of the thermo-magnetic dependence in the coupling constant of the model. It can be seen from the subsequent results presented in this mini-review that the temperature and magnetic field dependent coupling constant of NJL model helps describing the actual physical phenomena around the transition temperature in presence of strong magnetic field and in the process improves the compatibility with the lattice results.

Finally, after revisiting most of the existing studies on the literature our main conclusion is that we still need a rigorous approach from the microscopic level for the betterment of understanding this beautiful and intriguing phenomena of inverse magnetic catalysis predicted by lattice simulations.

{\it Acknowledgement } This work was partially supported by Conselho Nacional de Desenvolvimento Cient\'{\i}fico e Tecnol\'ogico (CNPq) under Grant No. 304758/2017-5 (R.L.S.F.), Funda\c{c}\~ao de Amparo \`a Pesquisa do Estado do Rio Grande do Sul - FAPERGS, Grants No.19/2551-0001948-3 (R.L.S.F.) and No. 19/2551-0000690-0 (R.L.S.F.). R.L.S.Farias is also grateful to Sidney A. Avancini, Marcus B. Pinto, Varese S. Tim\'oteo, Gast\~ao Krein, William R. Tavares, Norberto Scoccola, Dyana C. Duarte and Rudnei O. Ramos for fruitful collaboration and numerous discussions over the years.  R.L.S.Farias also benefited from discussions and would like to acknowledge Jens Andersen,  Alejandro Ayala and Gergo Endr\"{o}di.

\end{document}